\documentclass[lettersize, journal]{IEEEtran}
\usepackage{array}
\usepackage{stfloats}
\usepackage{url}
\usepackage{verbatim}

\usepackage{cite}
\usepackage{amsmath,amssymb,amsfonts}
\usepackage{graphicx}
\usepackage{textcomp}
\usepackage{xcolor}
\usepackage{epstopdf}
\usepackage{lineno}
\usepackage{algpseudocode}
\usepackage{algorithmicx,algorithm}
\usepackage{subfigure}
\usepackage{booktabs}
\usepackage{multirow}
\usepackage{makecell}
\usepackage[T1]{fontenc}
\usepackage[utf8]{inputenc}
\usepackage{authblk}
\usepackage{amsthm}
\usepackage{bm}
\usepackage{diagbox}
\usepackage{setspace}
\usepackage{hyperref}
\usepackage{xcolor}

\begin{document}
\title{Integrated Sensing, Communication, and Computing: An Information-oriented Resource Transaction Mechanism}
\author{Ning~Chen,~\IEEEmembership{}
       Zhipeng~Cheng,~\IEEEmembership{}
       Xuwei~Fan,~\IEEEmembership{}
       Zhang~Liu,~\IEEEmembership{}
       Bangzhen~Huang,~\IEEEmembership{}
       Jie~Yang,~\IEEEmembership{}
       Yifeng~Zhao,~\IEEEmembership{}
       Lianfen~Huang ~\IEEEmembership{}\\
\thanks{Corresponding author: Yifeng Zhao (e-mail:zhaoyf@xmu.edu.cn)}
\thanks{Ning Chen, Xuwei Fan, Zhang Liu, Bangzhen Huang, Yifeng Zhao, and Lianfen Huang are with the Department of Information and Communication Engineering, Xiamen University, 361005 Xiamen, China.}
\thanks{Zhipeng Cheng is with the School of Future Science and Engineering, Soochow University, 215006 Suzhou, China.}
\thanks{Jie Yang is with the National Institute for Data Science in Health and Medicine, Xiamen University, 361102 Xiamen, China.}
%


}



\maketitle

\begin{abstract}
Information acquisition from target perception represents the key enabling technology of the Internet of Automatic Vehicles (IoAV), which is essential for the decision-making and control operation of connected automatic vehicles (CAVs). Exploring target information involves multiple operations on data, e.g., wireless sensing (for data acquisition), communication (for data transmission), and computing (for data analysis), which all rely on the consumption of time-space-frequency-computing (TSFC) multi-domain resources. Due to the coupled resource sharing of sensing, communication, and computing procedures, the resource management of information-oriented IoAV is commonly formulated as a non-convex NP-hard problem. In this article, further combining the integrated sensing and communication (ISAC) and computing, we introduce the integrated sensing, communication, and computing (ISCC), wherein the TSFC resources are decoupled from the specific processes and shared universally among sensing, communication, and computing processes. Furthermore, the information-oriented resource trading platform (IRTP) is established, which transforms the problem of ISCC resource management into a resource-information substitution model. Finally, we embed the employment topology structure in IoAV into neural network architecture, taking advantage of the graph neural network (GNN) and multi-worker reinforcement learning, and propose the dynamic resource management strategy based on the asynchronous advantage GNN (A2GNN) algorithm, which can achieve the convergence both of information gain maximization and resource consumption minimization, realizing efficient information-oriented resource management.
\end{abstract}


\section{Introduction}
\IEEEPARstart{I}{nternet} of Automatic Vehicles (IoAV) is an advanced cross application derived from the intersection of the Internet of Vehicles (IoV) (e.g. C-V2X) and automatic driving. IoV endows autonomous driving with the ability of information interaction and resource sharing, which makes up for the deficiency of single-vehicle automatic driving, such as the insufficient ability of sensing and decision-making timeout caused by limited computing resources \cite{b1, b2}. The control decisions of automatic vehicles (CAVs) depend on the obtained target information, while information cannot exist independently, and data is the carrier of information. Consequently, aiming to mine the target information, CAVs need to collect the sensing data containing the information, and then analyze and extract the target information utilizing the local computing resources and the edge computing resources deployed in road side units (RSUs). Note that CAVs need to unload part or all of the sensing data to RSUs drawing support from the communication transmission if they desire to utilize the computing resources in RSUs. Accordingly, facing the large-scale sensing data generated by CAVs, to achieve a complete closed-loop design of the acquisition of target information, integrated sensing, communication, and computing (ISCC) is introduced in this article.

Integrated sensing and communication (ISAC) represents a key candidate technology for the 6G radio access network (RAN), which has attracted extensive attention and discussion recently \cite{b3, b4, b5}. ISAC can provide technical support for IoAV via making use of time, spectrum, and space resources at the sensing and communication level. On the one hand, wireless sensing, a fire-new sensing pattern besides the traditional sensing methods such as vision, radar, and lidar, is offered to CAVs, which contributes to detection, estimation, and recognition of targets \cite{b3}. On the other hand, ISAC builds high-speed communication links based on mmWave and large-scale antenna arrays, creating conditions for task offloading and information distribution between CAVs and RSUs \cite{b6}. Obviously, ISAC, focusing on wireless sensing and communication, puts less emphasis on the calculation and processing of the obtained sensing data, which misses a key link of target information acquisition.

\begin{table*}[htbp]
\center
\footnotesize
\caption{The foundation of sensing, communication, and computing}
\label{table_1}
\begin{tabular}{|c|c|c|c|c|}
\hline
              & Resource requirements                                                                 & Evaluation criteria                                                                           & Operation on data & Operation on information                                                            \\ \hline
Sensing       & \begin{tabular}[c]{@{}c@{}}Time \\ Space\\ Frequency\\ Energy\\ ...\end{tabular} & \begin{tabular}[c]{@{}c@{}}Detection probability\\ False-alarm probability\\ ...\end{tabular} & Data generation               & \begin{tabular}[c]{@{}c@{}}Initial information generation \end{tabular} \\ \hline
Communication & \begin{tabular}[c]{@{}c@{}}Time \\ Space\\ Frequency\\ Energy\\ ...\end{tabular} & \begin{tabular}[c]{@{}c@{}}Data rate\\ Error Rate\\ Reliability\\ ...\end{tabular}            & Data transmission               & Information Distribution
 \\ \hline
Computing     & \begin{tabular}[c]{@{}c@{}}Time \\ Computing Power\\ Energy\\ ...\end{tabular} & \begin{tabular}[c]{@{}c@{}}Complexity\\ Queuing delay\\ ...\end{tabular}                      & Data mining                  & Information Extraction  \\ \hline
\end{tabular}
\end{table*}

ISAC is the two-dimensional coordination of wireless sensing and communication. In more dimensions, Wei et al. studied the integrated wireless sensing, communication, and security, wherein the waveform design considering physical layer security is received attention \cite{b7}. Carlos et al. studied the integration of wireless sensing, communication, and location \cite{b8}. ISCC is the coordination of ISAC and computing, which further evolves from the two-dimensional combination of wireless sensing and communication to the three-dimensional cooperation of wireless sensing, communication, and computing. Consequently, the system complexity is further increased, especially for the resource management among the multiple operations of ISCC. It is a gordian knot that how to realize the efficient utilization of time-space-frequency-computing (TSFC) multi-domain resources on the premise of meeting the information requirements of user equipment (UE). Generally speaking, the resource allocation (RA) of ISAC is a non-convex NP-hard problem, which will also be obviously exposed to the RA in the ISCC. Meanwhile, the RA of ISCC shows new complex characteristics due to the consideration of computing \cite{b3, b4, b9, b9-1}. Table \ref{table_1} shows the relevant characteristics of wireless sensing, communication, and computing. Major difficulties of resource management in ISCC are analyzed below:

\begin{itemize}
  \item [1)] Internal association and coupling. Sensing, communication, and computing are sub-process during information acquisition, where their performance can affect the final information revenue. Besides, they are related and coupled, and each operation will affect others. For instance, the channel information extracted from wireless sensing and computing can reduce the signaling overhead of communication. Simultaneously, the calculation and analysis of the echo signal of communication can realize wireless sensing at almost zero additional spectrum and beam cost \cite{b6, b7}. In the information-oriented ISCC system, it is arduous to explore an independent optimization method that completely decouples wireless sensing, communication, and computing.

  \item [2)] Various performance evaluation criteria. The performance characterizations of wireless sensing, communication, and computing are diverse. First, wireless sensing mainly focuses on detection probability and false alarm probability \cite{b3, b4}. Then, communication emphasises more on data rate and transmission reliability \cite{b2, b6}. Finally, complexity and queuing delay represent the main focuses of computing \cite{b10}. Regarding the joint optimization of wireless sensing, communication, and computing, it faces difficulties to achieve the optimal performance indexes simultaneously, whereas the optimization of partial parameters cannot represent the overall performance. There is a lack of global performance evaluation standards in the resource management of the ISCC.

 \item [3)] Resource sharing and competition. The demand for TSFC multi-domain resources in wireless sensing, communication, and computing is not independent, i.e., overlap in resource consumption. As an example, time, space, and frequency are required by both wireless sensing and communication, simultaneously computing also needs time resources. The competition among wireless sensing, communication, and computing in TSFC multi-domain resources pose great challenges to evaluate the overall resource efficiency only by a single process, which further complicates the resource optimization for a separate program in ISCC \cite{b2, b6}.
\end{itemize}


\begin{figure*}[t]
	\centering
	\includegraphics[width=6.0in]{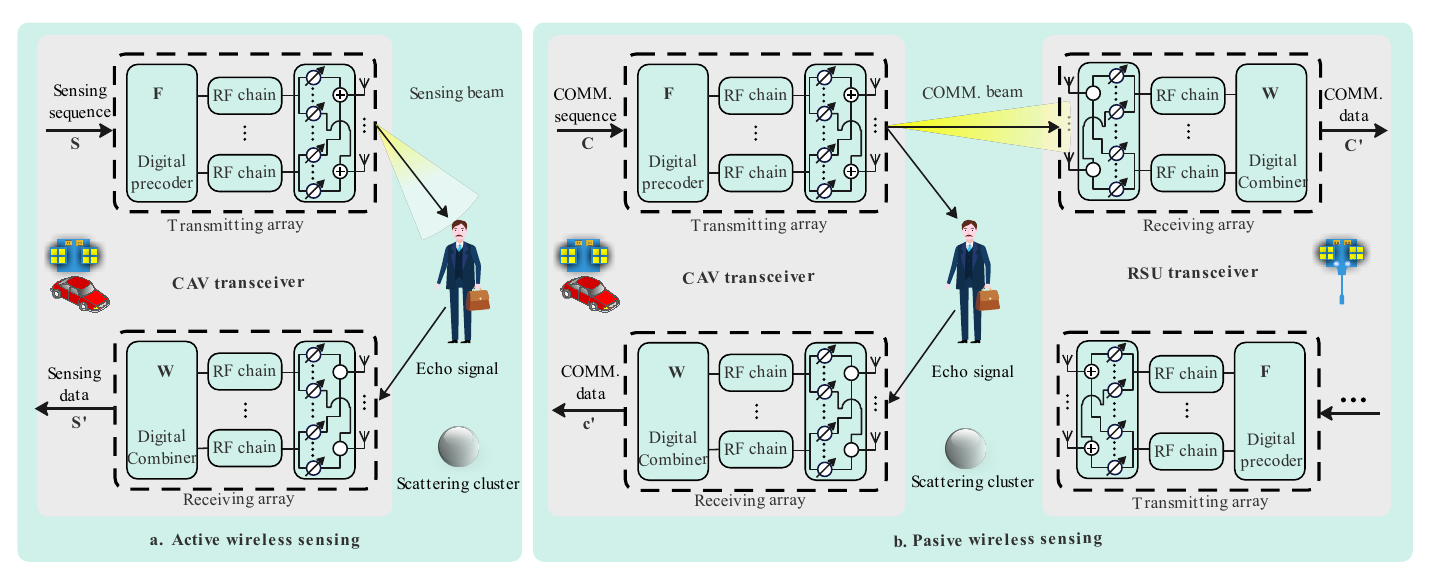}
    \caption{The active wireless sensing and passive wireless sensing.}
	\label{fig_1}
\end{figure*}

To this end, it is crucial to design an evaluation target that represents the overall performance of  ISCC system involving wireless sensing, communication, and computing. On this basis, developing an efficient TSFC resource management algorithm that determines the application prospect of the ISCC is critical and urgent. To tackle the above-mentioned problems, we propose an information-oriented resource transaction mechanism. Major contributions are summarized as follows.

\begin{itemize}
  \item [1)] TSFC multi-domain resources are decoupled from wireless sensing, communication, and computing, which releases the binding of resources and procedures, and establishes the common twin resource pools, realizing deterministic resource scheduling during the information acquisition.
  \item [2)] Following the essence of the relationship between employers and employees, an information-oriented resource trading platform (IRTP) is built in this work, wherein the substitution pattern between TSFC multi-domain resources and information is formed, which converts the complex resource management into an interpretable market transaction model.
 \item [3)] We integrate the employment topology structure among CAVs into the neural network (NN) structure and further combine that with the reinforcement learning (RL) model with multi-worker, and propose the asynchronous advantage GNN (A2GNN) algorithm, which can realize the bidirectional optimization of the information gain and resource cost.
\end{itemize}

The rest of this paper is organized as follows. We review the foundation of ISAC and discuss the development prospect of ISCC in information-oriented IoAV in Section \uppercase\expandafter{\romannumeral2}. The information-oriented resource transaction model and the dynamic resource management strategy based on the A2GNN algorithmare presented in Section \uppercase\expandafter{\romannumeral3}. Finally, we conclude this paper in Section \uppercase\expandafter{\romannumeral4}.

\section{From ISAC to ISCC, the inevitable expansion of information-oriented network}
\subsection{ISAC}
ISAC refers to the unified design of wireless sensing and communication by the joint design of air interface and protocol, reuse of time-frequency-space resources, and the sharing of hardware equipment so that the wireless network can realize high-precision wireless sensing while carrying out high-quality communication transmission, which can improve the overall performance and service capability of the network. During the evolution of ISAC, the coexistence of radar and communication transceiver is the early development goal, wherein the spectrum sharing of wireless communication and radar sensing is the main research direction, including spectrum sensing, dynamic spectrum access and interference mitigation. Ulteriorly, ISAC is expected to realize the joint design of wireless sensing and communication on the hardware platform and network architecture \cite{b7}. On one hand, the integration of radar systems and communication systems realizes efficient spectrum utilization and low hardware cost. On the other hand, the coordination of wireless sensing and communication can promote each other and realize a win-win, such as communication-assisted sensing and sensing-assisted communication.

%

\begin{table*}[htbp]
\caption{Comparison between active wireless sensing and passive wireless sensing }
\label{table_2}
\center
\begin{tabular}{|c|c|c|c|c|c|}
\hline
    & Resource cost & Tracking & Stability & Flexibility & Region       \\ \hline
Active wireless sensing & High          & Yes      & High      & High        & Unrestricted \\ \hline
Passive wireless sensing & Low           & No       & Low       & Low         & Restricted   \\ \hline
\end{tabular}
\end{table*}

In the ISAC based on large-scale antenna arrays, wireless sensing can be divided into active wireless sensing (AWS) and passive wireless sensing (PWS), as shown in Fig. \ref{fig_1} \cite{b3}. In the AWS, the transmitting antenna arrays of the perceiver are required to transmit the mmWave beam carrying the special sensing sequence to the direction related to the target. Then the receiver arrays receive the echo signal reflected or scattered from the target, and then extract the interested target information contained in the data by computing. AWS needs to monopolize part of the spectrum and beam resources, but it is a region-unrestricted sensing mode that can flexibly adjust the frequency band and beam direction following the target. Conversely, in PWS, the perceiver does not need to transmit the sensing sequence to the target but uses the echo signal reflected or scattered by the communication beam or its multipath component on the target to obtain the target information in the communication with RSU. PWS does not need to allocate specific spectrum and beam resources for sensing, but it can only follow the communication direction, which belongs to the region-restricted sensing mode. Table \ref{table_2} shows the comparison between AWS and PWS.

Nowadays the research associated with ISAC can be divided into three directions: sensing-centric, communication-centric, and the trade-off between sensing and communication \cite{b4, b5}. However, from the perspective of information acquisition, the joint design and optimization of wireless sensing and communication is still still remains locally, which does not constitute a complete closed loop, and there is no overall planning for wireless sensing, communication, and computing for the search of target information. Meanwhile, the computational complexity of wireless sensing data analysis and processing, as well as the unloading and collaborative computing in resource-constrained scenarios are not taken into consideration\cite{b9, b9-1}. Therefore, ISCC, the further deep integration of ISAC and computing considering wireless sensing, communication, and computing from the overall process of target information acquisition, is of great significance to improve the performance limit of ISAC and the joint scheduling of TSFC resources.

\subsection{From ISAC to ISCC: a closed loop of information acquisition}

\begin{figure*}[t]
	\centering
	\includegraphics[width=6.0in]{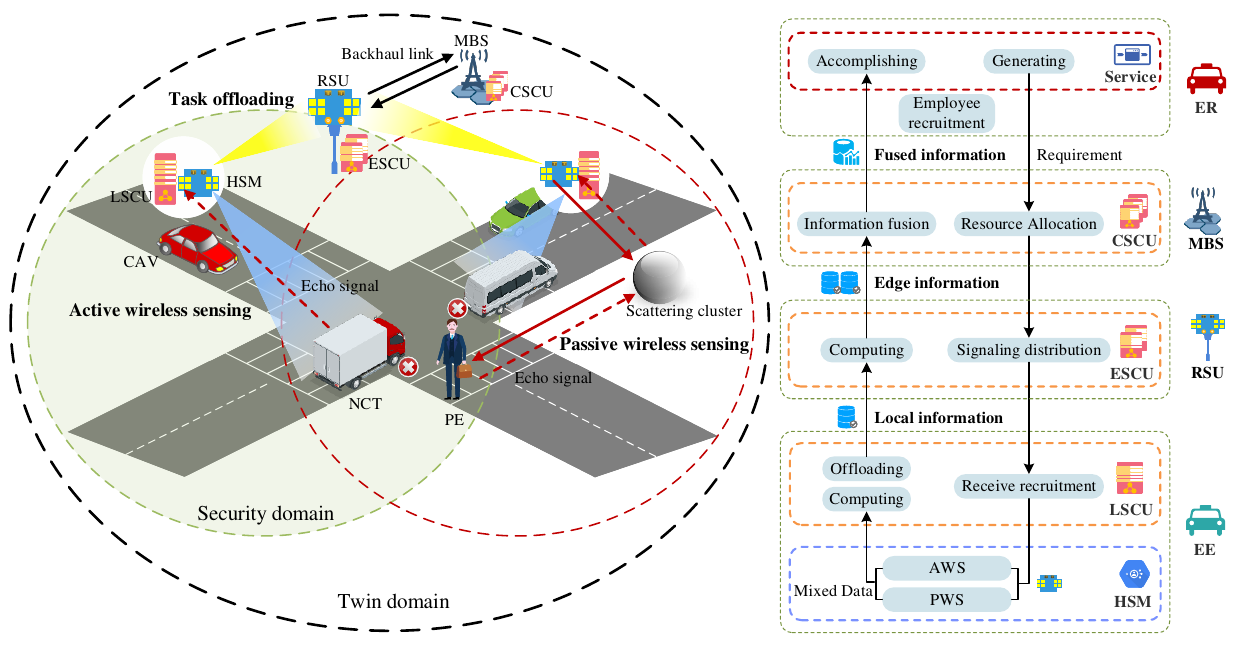}
    \caption{The information-oriented ISCC network model. CAVs with information requirements cooperatively implement wireless sensing based on large-scale MIMO, including active wireless sensing and passive wireless sensing. The original sensing data, which includes active wireless sensing data and passive wireless sensing data, can be calculated locally or unloaded to the RSUs. Finally, the CAVs pick up interested information from the mixed information.}
	\label{fig_2}
\end{figure*}

ISCC is the information-oriented closed loop of the IoAV, in which the data containing target information is generated by wireless sensing, and then offloaded via communication procedure, finally, computing is adopted for the extraction of information of interest in the sensing data. We consider an information-oriented IoAV network, as shown in Fig. \ref{fig_2}. The area that macro base station (MBS) can effectively cover is named twin domain (TD), wherein the resources can be flexibly scheduled based on the cell-free (CF) model, and centrally managed and controlled by the technologies of software-defined network (SDN), digital twin, and so on. It is assumed that a CAV can obtain the position and shape information of fixed objects such as roads and buildings through the high-precision map. At the same time, real-time information sharing can be realized among the networking equipment, such as CAVs, RSUs, and MBS in TD.

The area in which CAV can achieve safe driving is named as security domain, wherein CAVs need to obtain the wireless sensing data related to non-connected targets (NCTs) (e.g. non-connected vehicles (NCV) and pedestrian (PE)) based on the hybrid sensing module (HSM) composed by AWS and PWS. Then, the information of interest is extracted from the sensing data by computing, and finally the corresponding vehicle control decisions such as deceleration and avoidance are carried out based on the obtained target information. Due to the limited vision and sensing capacity, CAVs can detect targets within the security domain with the cooperation of multiple vehicles. Meanwhile, subject to the limited computing resources of the local storage and computing unit (LSCU), CAVs can transmit the wireless sensing data to the RSU by the mmWave communication of ISAC, and extract the information in the sensing data efficiently by the edge storage and computing unit (ESCU) with richer computing resources. The final information for CAVs is obtained by the fusion of local information and cloud information.

In the ISCC-based IoAV, the number of CAVs changes dynamically, while the driving environment and the information of NCTs to be obtained also change rapidly. In addition, NCTs have mobility controlled by their subjective consciousness, resulting in their number and location changing dynamically, which is an uncontrollable burst that cannot be controlled and managed by IoAV. Therefore, the network topology and collaborative relationship of information-oriented IoAV are time-varying. Meanwhile, the TSFC resources that ISCC requires are also dynamic, and constraints are different from times, such as scenes with exacting delay constraints, compared with other resources, it can be considered that time resources are seriously limited. With a view to the whole ISCC, how to realize an efficient application of TSFC resources while ensuring CAVs obtain sufficient target information which can ensure safe driving of vehicles, that is, to use lower resource cost to obtain higher information rate, is a significant and complex problem.

\section{Information-oriented Resource Transaction Mechanism Based on Graph model}
\subsection{Information-oriented Resource Trading Platform}

\begin{figure*}[t]
	\centering
	\includegraphics[width=6.0in]{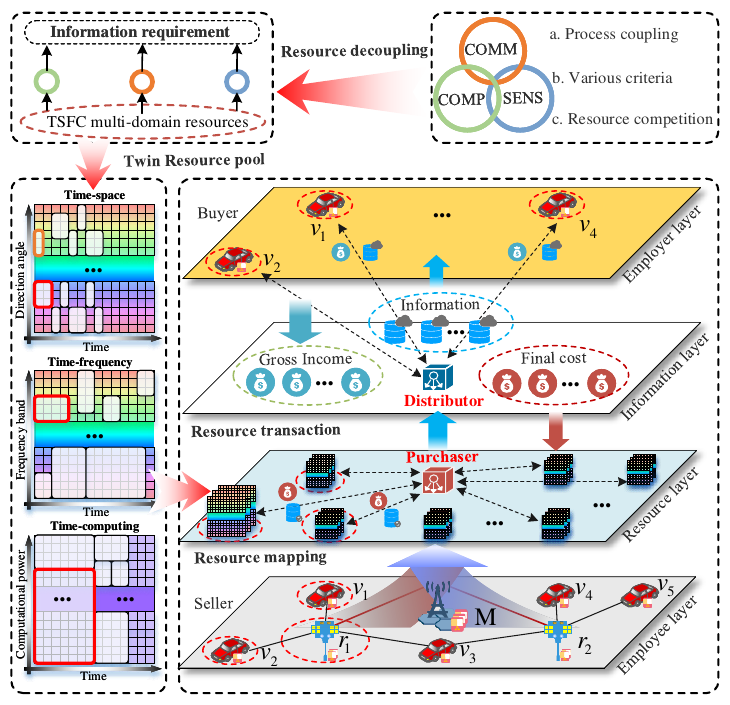}
    \caption{The construction of twin resource pool and the information-oriented resource trading platform. TSFC multi-domain resources are decoupled from the specific procedures in ISCC, and converted to three discrete two-dimensional resources.  The employment relationship is transformed into that the employer purchases information from the employee, and the employee uses his own resources to produce information for the employer. Middlemen "distributor" and "purchaser" optimize the transaction from the perspective of information and resources respectively.}
	\label{fig_3}
\end{figure*}

\begin{figure*}[htbp]
	\centering
	\includegraphics[width=5.0in]{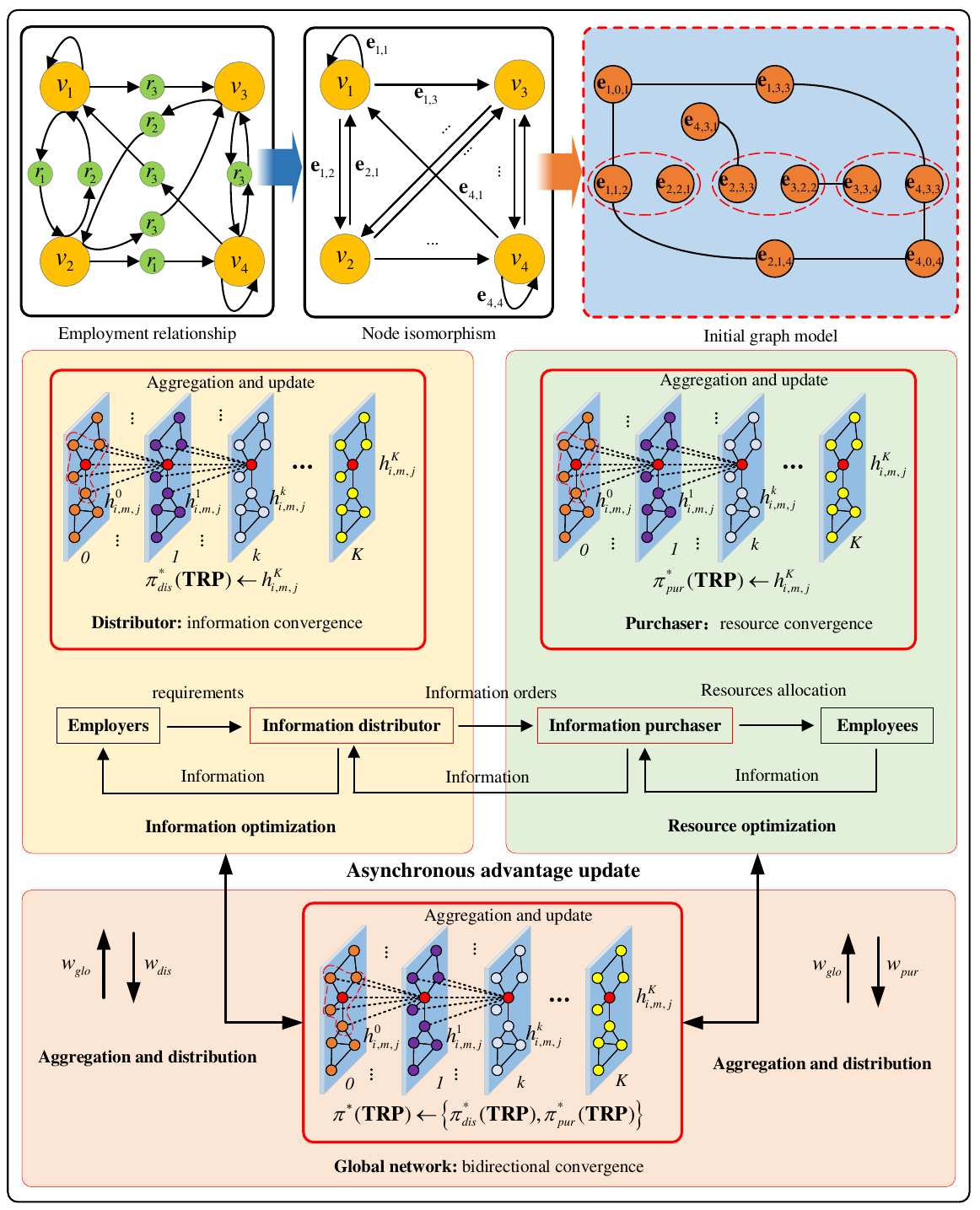}
    \caption{The resource transaction strategy based on graph model.}
	\label{fig_4}
\end{figure*}

TSFC resources required for ISCC are shared among the sensing, communication, and computing processes, resulting in multiple attributes of the same resource. To address this problem, we propose a twin resource pool (TRP) model that decouples TSFC resources from sensing, communication, and computing, that is, TSFC physical resources are mapped into logical resources based on advanced technologies, e.g., software-defined network (SDN) and digital twins. TRP is not exclusive to any one of sensing, communication, and computing, while it is universally shared by the ISCC. In particular, since the resources in the spatial domain, frequency domain, and computing power domain all have time attributes, we integrate time resources into the other three in TSFC. Consequently, TRP is composed of time-space resources, time-frequency resources, and time-computing resources, whose ordinates mean the direction angle, frequency band (i.e. subcarrier) and computing power respectively, as shown in Fig. \ref{fig_3}. Based on the above programs, the TSFC multi-domain resource management problem of IoAV is transformed into the problem of how to efficiently make use of the above three two-dimensional resources in TRP in the overall process of obtaining target information.

Based on decoupling TRPs, simulating the substitution relationship between information and resources from the perspective of the market economy, we propose the information-oriented resource trading platform (IRTP), as shown in Fig. \ref{fig_3}, where the networked and collaborative IoAV is modeled as the employment network of CAVs, and the employer reasonably recruits the employee (i.e. the connectable CAVs) via the connected RSUs to cooperatively implement the acquisition of NCT information in the security domain. Especially, employers who employ other CAVs may perform information acquisition tasks for other employers as employees simultaneously.

The employment relationship in IRTP means that the employer-CAV purchases the information produced by employee-CAVs with a certain amount of payment, while the employee-CAVs consume their own resources in TRP as the raw materials to produce information. Therefore, in essence, the transaction in IoAV where information is considered as a product means that employers purchase employees' resources in TRPs according to information requirements. Note that the information about all NCTs in the security domain is expected to be obtained accurately. Nevertheless, before obtaining information about specific NCTs, employers are broad and unclear about the information they need. Therefore, employers are vague about the information they should buy at first, and we can't formulate a resource trading strategy following the employer's information demand, which likes looking for a needle in a haystack. To address this problem, we propose the concepts of information purchaser and information distributor, wherein the information demand-oriented transaction network is virtually converted into two layers, the information production layer and the information sales layer, which undertake different marketable sub-tasks respectively.

As shown in Fig. \ref{fig_3}, the distributor orders information from the purchaser while the purchaser purchases information from the scattered resource sellers (i.e. employees) following the distributor's information order. Then purchaser feeds back the collected information to the distributor, and the distributor peddles information to the buyers (i.e. employers) who only select the interesting information to purchase and pay based on the amount of information available. By the above transactions, the information demand-oriented transaction is transformed into pre-customized information production and sales. First, the purchaser's optimization task is to reduce the final cost of purchasing sellers’ resources required in information production, which means to complete the distributor's information production task with less resource consumption, that is, to provide higher TSFC resource efficiency as far as possible. Then, the optimization task of the distributor is to formulate an information pre-purchase strategy that meets the market demand, so that the information purchased from the purchaser can get more gross income from the information buyers, improving the market hit rate and reuse rate of the pre-purchased information.

To sum up, the purchaser mainly improves efficiency from the perspective of resources, while distributors focus on the amount of the obtained useful information from the perspective of information. Under the constraint of meeting the information requirements of buyers and the resources in TRP of sellers, the net profits, which is the difference between the gross income and the final cost, can be regarded as the gain of the TSFC resource management in the IoAV, and the greater the value, the better the performance on the substitution of resources and information.

\subsection{Dynamic resource transaction based on graph model}

In the previous subsection, we modeled the NCT perception problem of CAVs as an information-oriented resource transaction model, in which the two middlemen "distributor" and "buyer" worked hard to increase the amount of effective information and reduce resource consumption respectively. Then, an efficient and reasonable method is required to guide the work of "distributor" and "purchaser", and effectively combine the breakthroughs achieved by the two to achieve efficient performance in terms of information and resources.

Machine learning (ML) is a promising tool in the resource allocation of wireless networks \cite{b11}. However, the architecture of neural networks used in the existing work has poor scalability and generalization and are lack of explicability. A method to improve scalability and generalization is to merge the structure of the target task into the neural network architecture \cite{b11, b12, b13}. In this paper, we propose an asynchronous advanced GNN (A2GNN) algorithm, in which the GNN model is used to embed the employment topology structure between CAVs into the neural network. Meanwhile, the multi-worker mechanism in reinforcement learning is adopted to enable the distributor and purchaser to learn interactively with the environment in different threads \cite{b14}. At the same time, workers’ parameters are uploaded, merged, and distributed with the global network. Finally, benefiting from the efforts of the two workers in different directions, the global network realizes the bidirectional convergence of information acquisition and resource consumption. The steps of A2GNN algorithm are as follows.

\begin{itemize}
  \item [1)] Generation of initial graph model. As shown in Fig. \ref{fig_4}, first, the employment relationship between employer and employee is modeled as a directed graphical model, wherein the employers, RSUs, and employees act as nodes, and the direction of the edges is the direction of the information flow in the employment relationship. In particular, if a CAV performs the NCT sensing task by itself, the start and end nodes of the edge are its own and no RSU is connected. Furthermore, to simplify the graph model, we integrate the RSUs into the features of edges, and establish a bidirectional digraph containing CAVs nodes only. Then, further extracting graph information, we model the employment relationship as nodes, and obtain the initial graph model. The vertex features include the information generated by the employment and the resources consumed in the employment. The edge between vertices and the edge features depend on its role in the subnet and optimization objectives.
  \item [2)] Bidirectional aggregation and update. After obtaining the initial graph model, we will expand it to a couple-GNN networks: distributor and purchaser. Then, multi-layer perceptrons (MLPs) are used to perform nodes aggregation and update. The edges of the distributor network are characterized by repeated information, while the characteristics of the edge in the purchaser network are the reused resources. Meanwhile, the distributor network and the purchaser network should regularly interact with the global network for parameters update, when the global network will guide the parameter adjustment of the two workers following the principle of maximizing its net profit.
 \item [3)] Convergence of global networks. After continuous iteration and parameter updating, global networks will achieve convergence in the two dimensions of information and resources, so as to guide the dynamic resource management of ISCC.
\end{itemize}

To sum up, the GNN can be used to extract the characteristics of CAVs' employment relationship, which can be fundamentally nested into the machine learning network architecture to improve the learning efficiency. Meanwhile, taking advantage of multi-workers' asynchronous advantage update of reinforcement learning, our global network can obtain the learning outcomes of two sub networks, so as to realize the optimization of information-oriented resource transaction towards the maximization of target information and the minimization of resource consumption. Finally, we can obtain the minimum resource consumption on the premise of ensuring the NCT information requirements of all CAVs, which realize the efficient management of TSFC multi-domain resources of ISCC in IoAV.

\section{Conclusion}
This article investigates the problem of TSFC multi-domain resource management in NCT information acquisition of CAVs. First, further integrating ISAC and computing, we introduce the ISCC architecture and decouple TSFC resources with specific procedures to obtain the universal TRP. Then, we establish IRTP to model the resource allocation of ISCC as a resource-information market transaction. Finally, the employment topology structure in IoAV is merged into the neural network architecture, and the A2GNN algorithm is proposed, where the graph model is used to express the employment relationship and resource transaction strategy between CAVs, while the multi-worker reinforcement learning model is used to realize the bidirectional optimization and convergence of information and resource efficiency, which provides an effective solution for the multi-domain resource allocation of ISCC.

\newpage

%
%
%
%

\vfill

\end{document}